\documentclass[]{aastex7}
\usepackage{appendix}
\usepackage{CJKutf8}
\usepackage{amsmath}
\usepackage{chemformula}
\usepackage{chngcntr}
 \usepackage[section]{placeins}

\accepted{11/14/2025}

\submitjournal{\psj}

\shorttitle{}
\shortauthors{}

\begin{document}
\begin{CJK*}{UTF8}{gbsn}

\title{Snowball Bistability Vanishes at Moderate Orbital Eccentricity}

\author[0000-0002-1592-7832]{Xuan Ji(纪璇)}
\affiliation{Department of the Geophysical Sciences, The University of Chicago, Chicago, IL 60637 USA}
\email{xuanji@uchicago.edu}

\author{Dorian S. Abbot}
\affiliation{Department of the Geophysical Sciences, The University of Chicago, Chicago, IL 60637 USA}
\email{abbot@uchicago.edu}

\begin{abstract}
Snowball episodes are associated with increases in atmospheric oxygen and the complexity of life on Earth, and they may be essential for the development of complex life on exoplanets. Sustained, stable Snowball episodes require a Snowball bifurcation and climate bistability between the globally ice-covered Snowball state and a state with at least some open ocean. We find that climate bistability disappears for an aquaplanet with a slab ocean in the global climate model ExoCAM when the orbital eccentricity is increased to 0.2-0.3. This happens because the Snowball state ceases to exist as seasonal insolation variations intensify, while the warm state remains stable due to the ocean's large heat capacity. We use a low-order ice-thermodynamic model to show that the Snowball state ceases to exist as seasonality increases because winter freezing at the ice bottom is reduced relative to summer melting at the ice top due to ice self-insulation. Combined with previous research showing that Snowball climate bistability diminishes for planets orbiting low-mass stars and ones with longer rotation periods, and that it disappears entirely for tidally locked planets, our work suggests that the Snowball climate bistability may not be as robust to planetary parameters as previously thought, representing one aspect of habitability more consistent with the rare Earth hypothesis than the Copernican principle.

\end{abstract}


\section{Introduction}\label{sec:intro}


``Snowball Earth" events are global glaciations that last millions of years \citep{kirschvink1992late,hoffman_neoproterozoic_1998}. There seem to have been two periods in Earth history where one or more Snowball events occurred in relatively quick succession \citep{hoffman2017snowball}. Isotopic evidence indicates that these Snowball events coincided with major increases in atmospheric \ch{O2} \citep{tajika_great_2019}, which suggests a possible causal relationship between Snowball glaciations and the rise of atmospheric oxygen \citep[e.g.,][]{kirschvink_paleoproterozoic_2000, kasting_what_2013, claire_biogeochemical_2006, harada_transition_2015}. The increase in \ch{O2} altered the redox state of Earth's surface environment and influenced life and its evolution. Furthermore, \ch{O2} is a bioindicator in the search for life on exoplanets \citep{harman_abiotic_2015, meadows_exoplanet_2018, krissansen-totton_oxygen_2021}. Together, these factors highlight the Snowball state as an important phase in planetary evolution and the development of life. In this context, understanding the prevalence of Snowball episodes on exoplanets could offer insight into the conditions required for complex life and thus inform the Copernican Principle or the Rare Earth Hypothesis \citep{ward_rare_2000}.

A crucial aspect of the Snowball Earth hypothesis is an extended period in the globally glaciated state \citep[supported by U-Pb dating, see][]{hoffman2017snowball}, which implicates all relevant features discussed above, such as changes in atmospheric \ch{O2} and pressure on life that might result in increased innovation. From a theoretical standpoint, nonlinearity caused by the albedo contrast between sea ice and open ocean allows for bifurcations, or tipping points, in planetary climate, as a Snowball is entered and exited, as well as hysteresis and bistability in planetary climate such that both the snowball and a less glaciated state exist for the same external forcing of, e.g., stellar flux and CO$_2$ \citep{budyko_effect_1969, sellers_global_1969}. Climate hysteresis is supported by cap carbonates overlying Snowball layers in geological history, suggesting that CO$_2$ had to build up to immense levels during Snowball events until equatorial ice could be melted, after which the CO$_2$ was deposited in the carbonates \citep{hoffman_neoproterozoic_1998}. If there were no hysteresis in planetary climate, a planet would be unlikely to stay in a Snowball long, since the weathering that removes CO$_2$ from the atmosphere would likely be greatly reduced, and CO$_2$ outgassing would quickly raise the temperature enough to cause deglaciation  \citep{menou2015climate,abbot2016analytical}. These climate bifurcations and bistability are robust aspects of Earth's climate that can consistently be produced by idealized models with different levels of complexity \citep[e.g.][]{budyko_effect_1969, sellers_global_1969, north1975analytical, ghil_climate_1976, caldeira_susceptibility_1992, lucarini_thermodynamic_2010, roe_notes_2010, abbot_jormungand_2011, boschi_bistability_2013, lewis_neoproterozoic_2003, donnadieu_impact_2004, deitrick_functionality_2023} and by global climate models (GCMs) \citep[e.g.][]{wetherald_effects_1975, marotzke_presentday_2007, vizcaino_long-term_2008, voigt_transition_2010, ferreira_climate_2011, wolf_constraints_2017, ramme_climate_2022, eisenman_radiative_2024, obase_climate_2025}, as reviewed in \cite{pierrehumbert_climate_2011} and \cite{von_storch_multiple_2025}.

Recent work suggests that climate bistability may not be as robust on all exoplanets as it is on Earth. For example, an M-star stellar spectrum weakens the ice-albedo feedback and reduces climate hysteresis \citep{joshi_suppression_2012, shields_effect_2013, shields_spectrum-driven_2014}. More strikingly, the Snowball bifurcation is likely to disappear entirely on slowly rotating \citep{lucarini_habitability_2013, abbot_decrease_2018} and tidally locked \citep{checlair_no_2017, checlair_no_2019-1} planets. Together, these effects make global Snowball episodes unlikely for potentially habitable planets orbiting M-stars.

\begin{figure}
    \centering
    \includegraphics[width=0.5\linewidth]{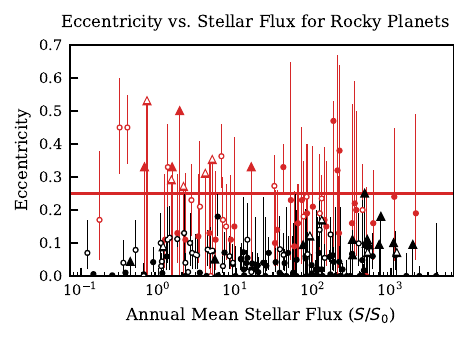}
    \caption{Eccentricity versus annual-mean stellar flux for rocky exoplanets. Dots represent measured eccentricities, with vertical lines showing observational uncertainties; triangles denote planets for which only upper-limit constraints on eccentricity are available. Red symbols highlight planets whose upper eccentricity limits exceed 0.25, while black symbols represent the rest. The annual-mean stellar flux is computed using the mean eccentricity via $S \propto (a^2 \sqrt{1 - e^2})^{-1}$. Only potential rocky exoplanets with constrained eccentricity measurements are included in the sample. 72 out of 216 rocky planets have eccentricity upper limits exceeding 0.25. The sample is drawn from the confirmed exoplanet catalog on the NASA Exoplanet Archive \citep{nasa_exoplanet_archive_confirmed_2019}: for transiting planets, the criteria for selecting ``rocky planets" follow \citet{ji_cosmic_2025} (filled symbols); for planets without radius measurements, those with planetary masses less than 6 Earth masses are shown (open symbols).}
    \label{fig:eccen}
\end{figure}

Eccentricity is another important parameter that could impact Snowball climate bistability. Multiple mechanisms could increase the eccentricities of rocky planets, including excitation by massive perturbers such as giant planets or stellar companions, as well as secular and resonant interactions \citep{spiegel_generalized_2010, lithwick_eccentric_2011, ida_toward_2013, Kane_2014}. When accounting for observational uncertainties, 33\% of rocky exoplanets have 1-$\sigma$ upper limits on eccentricity that exceed 0.25 (Fig.~\ref{fig:eccen}), much larger than Earth's value of 0.0167. At higher eccentricities, it is harder to stabilize perennial ice at the poles or equator \citep{wilhelm_ice_2022, deitrick_exo-milankovitch_2018}. Moreover, \citet{linsenmeier_climate_2015} found a large reduction in Snowball climate hysteresis and bistability when the eccentricity is increased from 0 to 0.5. They used PlaSim, a general circulation model of intermediate complexity that incorporates a 0-layer Semtner ice scheme, which is insufficient to capture the diurnal surface temperature cycle of equatorial ice, which is crucial for Snowball deglaciation \citep{abbot2010importance}. Taken together, this body of work motivates a more thorough investigation of the impact of orbital eccentricity on Snowball climate bistability.

In this study, we show that Snowball climate bistability vanishes if the eccentricity is increased to 0.2--0.3 in the sophisticated GCM ExoCAM \citep{wolf_exocam_2022}. We then interpret and explain our results using a low-order ice thermodynamic model that was developed to understand seasonal variations in Arctic sea ice and associated bifurcations \citep{eisenman_nonlinear_2009}. The loss of climate bistability is primarily driven by the loss of stability of the Snowball state as the eccentricity increases. Ice self-insulation reduces freezing at the bottom of the ice relative to melting at the top, such that increased seasonality causes more melting than freezing and makes the Snowball state less stable.




%

The plan of this paper is as follows. In section~\ref{sec:model}, we introduce the GCM  and low-order ice–thermodynamic model. In section~\ref{sec:results}, we present the GCM results showing that Snowball bistability disappears at eccentricities greater than 0.25, driven by summer melting at periastron. We also explore how this transition depends on key parameters. We discuss our results in section \ref{sec:discuss} and conclude in section \ref{sec:conclusion}.

\section{methods}\label{sec:model}

\subsection{Global Climate Model}

We use the 3D GCM ExoCAM\footnote{\url{https://github.com/storyofthewolf/ExoCAM}} \citep{wolf_exocam_2022}, which is designed to be flexible for exoplanet applications. ExoCAM  has been used widely to study the climate and atmospheric circulation of exoplanets \citep[e.g.][]{wolf_hospitable_2013, wolf_delayed_2014, wolf_controls_2014, yang_differences_2016, kopparapu_habitable_2017, haqq-misra_demarcating_2018, badhan_stellar_2019, kang_regime_2019, komacek_atmospheric_2019, komacek_scaling_2019, yang_effects_2019, adams_aquaplanet_2019, chen_persistence_2020, fauchez_trappist-1_2020, hu_o2_2020, rushby_effect_2020, suissa_first_2020, zhang_evolving_2021, li_rotation_2022, kossakowski_carmenes_2023, rotman_general_2023, wang_lorenz_2023, lobo_terminator_2023, hammond_coupled_2024,liu_hydrologic_2024, Zhan2024}. We use an aquaplanet surface (no continents) and a 50-m slab (mixed layer) ocean with zero imposed ocean heat transport. We consider eccentricities up to 0.3, at which the native ExoCAM code for calculating the true longitude (anomaly) is inaccurate. We therefore modified the ExoCAM orbital module to replace its polynomial approximation with a numerical solver of Kepler’s equation, following the method of \citet{adams_aquaplanet_2019}. We also updated the calendar module to allow the orbital period to be set to any value and to ensure it remains coupled with the orbital module. These changes were applied to all simulations in this study. We run the model at f45 horizontal resolution (finite volume $4^\circ \times 5 ^\circ$) with 40 vertical levels. We set the dry atmospheric surface pressure to 1 bar, with 400 ppm \ch{CO2} and the remainder \ch{N2}. All simulations use the Sun's stellar spectrum and a planetary rotational period of 24 hours. As we vary the eccentricity, we keep the annual-mean stellar flux constant and fix the orbital period to 360 days. Unless otherwise noted, we set the obliquity to zero.

Our approach for investigating bistability is as follows. First, we perform simulations at stellar fluxes of $1600\ \text{W}/\text{m}^2$, which results in a climate without sea ice, and $500\ \text{W}/\text{m}^2$, which results in a completely ice-covered ocean. For each parameter set, we initiate simulations from both ice-free conditions (hereafter Hot Start) and ice-covered conditions (hereafter Cold Start). The model is bistable for a set of parameters if the Hot Start and Cold Start simulations equilibrate to different climates. We evaluate simulation equilibrium using the metrics that the absolute value of the sum of the top-of-atmosphere (TOA) and surface energy imbalance is less than $1\ \text{W}/\text{m}^2$ and the year-on-year change in surface temperature in the last five years is less than 0.2 K/yr. We average all relevant variables over five years of simulation after equilibrium has been reached. 




\subsection{Low-order Ice-thermodynamic Model}\label{sec:ice-model}

In order to better understand our GCM results and consider the effects of varying uncertain parameters, we use a low-order ice thermodynamic model originally proposed by \citet{eisenman_nonlinear_2009} and later used in a number of studies of Arctic sea ice stability and bifurcations \citep{abbot_bifurcations_2011, eisenman_factors_2012, wagner_how_2015, hill_analysis_2016}. We make particular use of the analytical insights of \citet{hill_analysis_2016} and adopt their assumption of no basal heating, which is consistent with our GCM setting. The \citet{eisenman_nonlinear_2009} model was originally developed for the sea ice annual cycle in the Arctic, and we adapt it to apply to the equatorial region of a Snowball climate state that experiences an annual cycle in stellar forcing due to an eccentric orbit. Our main purpose will be to investigate the destabilization of the Snowball state that leads to the transition to a non-Snowball climate.


\begin{table}[h]
\centering
\begin{tabular}{c|c|c|c}
Parameter&Symbol&Value&Unit\\
\hline
\hline
Stellar flux at 1 AU &$S_0$&$1360$ & $\text{W/m}^2$\\
Latent heat of fusion of ice &$L_i$&3.00$\times 10^8$&J$\cdot$m$^{-3}$\\
Ice thermal conductivity&$k_i$&2.00&W$\cdot$m$^{-1}\cdot$K$^{-1}$\\
OLR constant&$A$&173 &W$\cdot$m$^{-2}$\\
OLR coefficient&$B$&1.44 &W$\cdot$m$^{-2}$$\cdot$K$^{-1}$\\
TOA albedo with ice-covered surface& $\alpha_{i}$ & 0.55 & -\\
Equatorial advective heat flux constant & $Q^*_{adv}$ & -22 & W$\cdot$m$^{-2}$\\
Equatorial advective heat flux coefficient & $\Delta Q_{adv}$ & -100 & W$\cdot$m$^{-2}$\\

\end{tabular}
\caption{Default parameters used in the low-order ice thermodynamic model.} 
\label{tab: para}
\end{table}

We now briefly review the model. For our application, we only consider the regime of a fully ice-covered state throughout the year, which allows us to simplify the model specification relative to \citet{eisenman_nonlinear_2009}. In this regime, the state variable is the thickness of sea ice, $h$, and we are solving for conditions that allow $h$ to reach zero at some point in the annual cycle, corresponding to a transition from the Snowball climate state to the non-Snowball climate state. $h$ has the following evolution equation,

\begin{align}
    \frac{dh(t)}{dt} & = -\frac{1}{L_i}({\text{Absorbed Stellar flux} + \text{Advective Flux} - \text{OLR}}) \\
    & =-\frac{(1 - \alpha_{i}) S_0}{\pi L_i r(t)^2} -\frac{Q_{adv}}{L_i} +\frac{A + B \cdot T_s(t)}{L_i} \\
    & =\frac{{B}\cdot T_{s} (t)-F(t)}{L_i},
    \label{eq:EBM}
\end{align}

where $F(t) \equiv \left[{(1 - \alpha_{i}) S_0}/({\pi r(t)^2})+Q_{adv}-A) \right]$ is the net absorbed energy—shortwave absorption plus advective energy (negative when heat is transported poleward from the equator)—minus the outgoing longwave radiation (OLR) that would be emitted with the ice top at the freezing temperature, assumed to be 0$^\circ$C. $r$ is the dimensionless distance from the planet to the star, scaled to one astronomical unit (AU), $T_s$ is the surface temperature in Celsius, and model parameters are defined in Table~\ref{tab: para}. We use a linear parameterization of outgoing longwave radiation (OLR) fit to output from our GCM simulations. To roughly fit the advective heat flux in the GCM, we scale it to annual mean stellar flux by letting $Q_{adv} = Q_{adv}^* + \left(S/S_0-1 \right) \cdot \Delta Q_{adv}$, where $S = S_0/\left(a^2\sqrt{1-e^2}\right)$, and $a$ is the semi-major axis of the planet scaled to one AU.

The eccentricity enters through the time dependence of the orbital distance $r$, which can be written as a function of the true anomaly ($\theta$): $r = {a(1-e^2)}/({1 - e\cos{\theta}})$. The time dependence of $\theta$ is obtained by solving Kepler’s equation.

In order to close the model, we need a specification of $T_s$. The maximum temperature the ice can reach is the freezing point $T_s=0$. When the net stellar input energy exceeds the sum of the OLR emitted by ice at the freezing temperature and the heat advected to higher latitudes ($F(t)>0$), the ice is melting, and $T_s = 0$. Otherwise  $T_s$ is determined by assuming surface energy balance with zero heat capacity: 

\begin{equation}
F(t)-B\cdot T_s -{k_i\cdot T_{s}}/{h} =0
\label{eq:Ts}
\end{equation}
where $k_i$ is the ice thermal conductivity. Rearranging, we can solve for surface temperature as $T_s = F(t)/B\cdot (h/(h+k_i/B))$. Then, following \cite{hill_analysis_2016}, the ice thickness evolution can be rewritten as:
\begin{equation}
\frac{dh}{dt} = \begin{cases}
-\frac{F(t)}{L_i}, & \text{ if } F(t) \geq 0 \\
\frac{-F(t)}{L_i}\cdot \left({1+h\frac{B}{k_i}}\right)^{-1}, & \text{ if } F(t) < 0
\end{cases}
\label{eq:EBM_simplify}
\end{equation}
The term $\left({1+h\frac{B}{k_i}}\right)^{-1}$ when $F(t) < 0$ is due to ice self-insulation. It results from the fact that ice freezes from the bottom, requiring heat to diffuse through the ice, which slows the process. There is no such term when $F(t) \geq 0$ because melting occurs at the ice surface under the assumption of zero basal heat flux in both the GCM and this model. This asymmetry will become important in our physical explanation and interpretation below.

We make a number of approximations in this model. For example, we approximate the heat capacity of ice as zero. Note that this assumption holds only for thin ice ($\lesssim 8 m$) \citep{eisenman_factors_2012}, which is consistent with our near-melting equatorial ice regime, though it would not hold for thick, global ice cover. We also approximate the heat capacity of the atmosphere as zero, assume a constant TOA albedo with ice present, and assume the freezing temperature of seawater is 0$^\circ$C. Nevertheless, we will find that the model fits GCM output reasonably well and yields useful qualitative insight and understanding.

Our low-order equatorial-ice model aims to capture and illustrate the mechanism that triggers the Snowball melt-out from the equator, rather than reproducing the full global hysteresis loop. Reproducing the critical global-mean ice fraction associated with Snowball initiation would require a one-dimensional model with spatial coupling between albedo and ice thermodynamics.

\section{Results}\label{sec:results}

\subsection{Loss of Snowball Climate Bistability at Moderate Eccentricity}
\label{sec:gcm-bistability}

\begin{figure}
    \centering
    \includegraphics[width=0.9\linewidth]{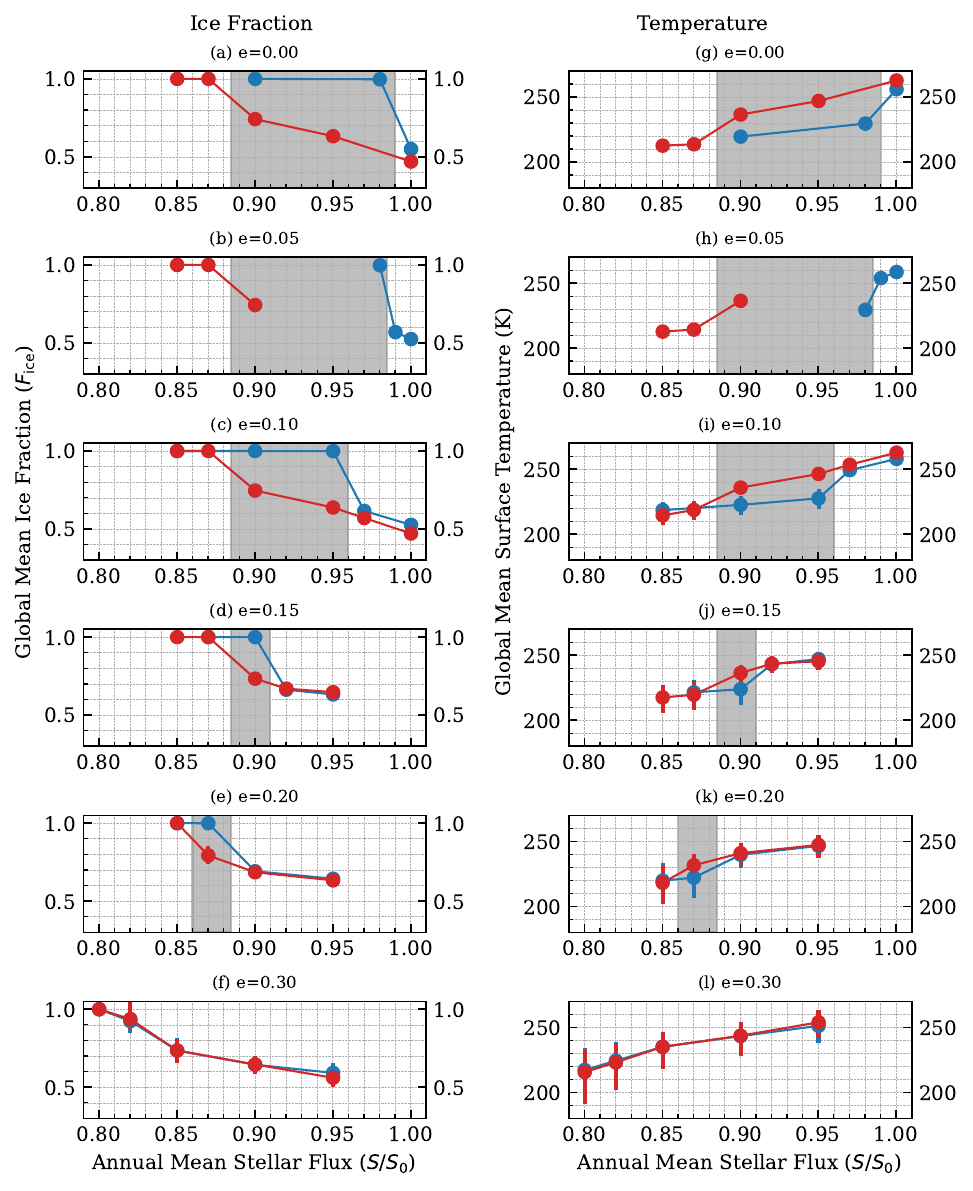}
    \caption{Snowball bifurcation diagrams for different eccentricities with the GCM ExoCAM. Blue circles represent simulations initialized with global ice coverage and red circles represent simulations initialized with no ice coverage. Seasonal variations in monthly-mean values are shown by vertical bars but are smaller than the circles in many cases. The gray shaded region indicates the range of stellar flux where Snowball bistability exists. The panels on the left ((a)-(f)) show the global-mean ice fraction (\( F_{\text{ice}} \)) as a function of annual-mean stellar flux. The panels on the right ((g)-(l)) show the global-mean surface temperature as a function of annual-mean stellar flux.}
    \label{fig:bifur-diagram}
\end{figure}


 

At small eccentricities we observe the standard situation with a wide region of climate bistability and bifurcations associated with transitions between climate states (Fig.~\ref{fig:bifur-diagram}). All non-Snowball states shown here exhibit an ice fraction greater than roughly 0.5, corresponding to an ice edge latitude near $30^\circ$. We refer to these climates, which feature a band of open ocean near the tropics, as Waterbelt states \citep{pierrehumbert_climate_2011}. The nonlinear transitions between Waterbelt and Snowball states are saddle-node bifurcations that involve an abrupt jump in climate. Snowball climate bistability requires both a Waterbelt-to-Snowball ($S^*_{WB\rightarrow SB}$) and a Snowball-to-Waterbelt ($S^*_{SB\rightarrow WB}$) bifurcation with a range of stellar fluxes between them where both the Snowball and Waterbelt climate states are possible for different initial conditions.

Climate bistability and bifurcations vanish as the eccentricity is increased from 0.2 to 0.3 (Fig.~\ref{fig:bifur-diagram}). Moderate eccentricity values can therefore destroy Snowball climate bistability, which was previously thought to be quite robust. The rest of this paper will be devoted to investigating and explaining this phenomenon.

The loss of climate bistability as the eccentricity increases is driven by a decrease in the stellar flux of the Snowball-to-Waterbelt bifurcation ($S^*_{SB\rightarrow WB}=0.99\ S_0$ at $e=0$ and $S^*_{SB\rightarrow WB}=0.89\ S_0$ at $e=0.2$). The Snowball-to-Waterbelt bifurcation occurs when seasonal ice melting exceeds seasonal ice freezing at the equator in the Snowball state. The decrease in $S^*_{SB\rightarrow WB}$ as eccentricity increases must therefore result from increased seasonality causing more summer melting than winter freezing. In section~\ref{sec:ice-thermo-implications} we will use the ice-thermodynamic model to argue that this is because melting occurs at the ice surface, whereas freezing occurs at the ocean-ice interface and is limited by the self-insulation of ice. In contrast, the stellar flux of the Waterbelt-to-Snowball bifurcation is nearly independent of eccentricity (\( 0.86\ S_0 < S^*_{WB\rightarrow SB} < 0.88\ S_0\)). This is because seasonal variations in surface temperature are small despite large variations in stellar flux in regions where there is open ocean due to the ocean's large heat capacity \citep[e.g.,][]{Williams2002, dressing_habitable_2010, bolmont_habitability_2016, Kane_2021, ji_inner_2023,liu_higher_2023, liu_eccentric_2024}. An increase in eccentricity therefore has relatively little effect on the tropical climate of a Waterbelt state and consequentially whether it freezes into a Snowball.

\subsection{Importance of Ice Thermodynamics}
\label{sec:ice-thermo-implications}

\begin{figure}
    \centering
    \includegraphics[width=1\linewidth]{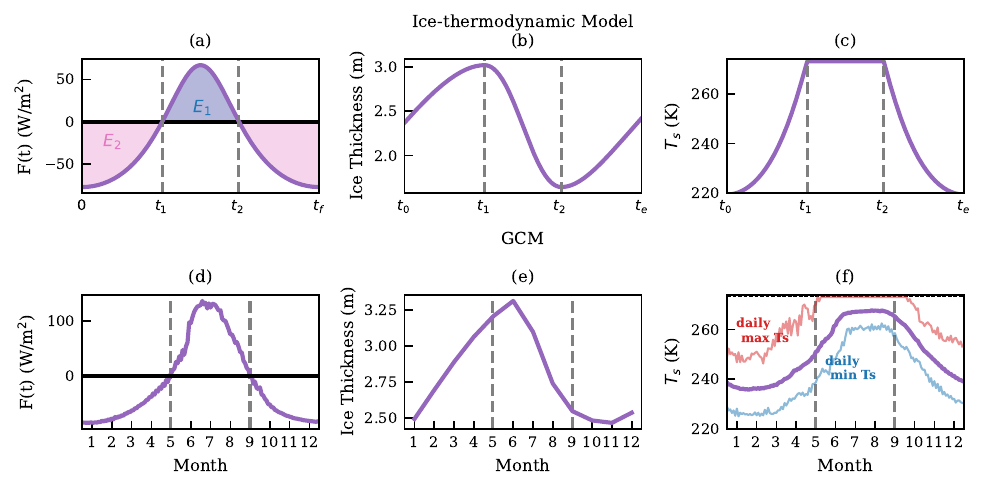}
    \caption{Comparison between equatorial ice thermodynamics model (top: panels a–c) and GCM output at the equator (bottom: panels d–f) near the limiting case where the Snowball state ceases to exist ($S/S_0$=0.87, e=0.2), demonstrating qualitatively similar behavior. The panels show $F(t)$ (Eq.~(\ref{eq:EBM})) (a,d), the ice thickness (b,e), and the surface temperature (c,f). $F(t)>0$ between times $t_1$ and $t_2$, leading to melting. We define the integral of $F(t)$ over this period as $E_1$, an important quantity discussed in the text. Similarly, freezing occurs when $F(t)<0$, and we define the integral of $F(t)$ over this period as $E_2$. The red and blue curves in panel (f) show the daily maximum and minimum temperature, while the thick purple line is the diurnal mean.}
    \label{fig:EBM_GCM}
\end{figure}

The low-order ice-thermodynamic model qualitatively captures the GCM's Snowball seasonal cycle in equatorial ice thickness well when we tune $\alpha_i$ and $Q_{adv}$ (Fig.~\ref{fig:EBM_GCM}). The main difference is that the temperature and ice thickness lag the forcing function by about 0.5-1 months in the GCM, but not in the ice-thermodynamic model because we assume zero heat capacity. Additionally, the ice-thermodynamic model does not include a diurnal cycle, but the GCM does. During summer ice melt, the daily maximum temperature in the GCM is the appropriate temperature for determining whether melting occurs \citep{abbot2010importance} and comparing with the ice-thermodynamic model temperature.  

Quantitatively, the magnitude of $F(t)$ differs between the low-order model and the GCM because both the albedo and advective heat flux vary seasonally. However, the annual-mean equatorial albedo and advective heat flux in the GCM are roughly independent of eccentricity, indicating that they are not the main drivers of melting with enhanced seasonality. We therefore keep them constant and tunable to better isolate the key mechanism. Further discussion is provided in Sec. \ref{sec:discuss}. 


Now that we have established that the ice-thermodynamic model provides a reasonable fit to the GCM results, we can use it to understand why increased seasonality reduces the ice thickness, all else being equal. During the ice melting phase ($F(t) \geq 0$) melting occurs at the top of the ice where heat is applied, such that heat goes directly into melting ice (Eq.~(\ref{eq:EBM_simplify})). During the ice growth phase ($F(t) < 0$) the situation is complicated by the fact that ice growth occurs at the bottom of the ice and the ice insulates itself, slowing growth. This can be seen from the extra 
factor of $\left({1+h\frac{B}{k_i}}\right)^{-1}$ in Eq.~(\ref{eq:EBM_simplify}) during the ice growth phase, which reduces $\frac{dh}{dt}$. As a result, increased seasonality (larger variation in $F(t)$) leads to more ice loss by melting than ice growth by freezing, and reduces the ice thickness, as we observed in the GCM in section~\ref{sec:gcm-bistability}.

We can use the ice-thermodynamic model to make quantitative predictions of the Snowball-to-Waterbelt transition by solving for the condition that the ice thickness reaches zero at its minimum. Following \citet{hill_analysis_2016}, we denote $t_1$ as the time at which $F(t) = 0$ with $\frac{dF}{dt} > 0$ and $t_2$ as the time at which $F(t) = 0$ with $\frac{dF}{dt} < 0$ (Fig.~\ref{fig:EBM_GCM} (a)). Assuming zero heat capacity, the ice thickness reaches its annual maximum, $h_m$, at $t_1$ and its minimum, which we will take to be $0$, at $t_2$. Integrating the ice thickness evolution equation (Eq.~(\ref{eq:EBM_simplify})) over the ice melting phase ($F(t) \geq 0$) yields,
\begin{eqnarray}
    \int_{t_1}^{t_2} \frac{dh}{dt}dt & = & \int_{t_1}^{t_2} -\frac{F(t)}{L_i}dt, \\
    \int_{h_m}^0dh & = & -\frac{1}{L_i} \int_{t_1}^{t_2} F(t)dt, \\
    h_m & =&  \frac{E_1}{L_i}, 
    \label{eq:ice-melt}
\end{eqnarray}
where we define $E_1 = \int_{t_1}^{t_2} F(t)dt$. Eq.~(\ref{eq:ice-melt}) says that the total ice melted during the melting phase is equal to the integral of the net heating of the ice over this period divided by the latent heat of melting. 

As mentioned above, during the ice growth phase ($F(t) < 0$) the situation is complicated by the fact that ice growth occurs at the bottom of the ice and the ice insulates itself, slowing growth. In this phase phase we can rewrite Eq.~(\ref{eq:EBM_simplify}) as
\begin{equation}
\left(1+h\frac{B}{k_i}\right) dh = \frac{-F(t)}{L_i}dt,
\end{equation}
which we can integrate over the ice growth phase to find
\begin{equation}
h_m+\frac{B}{2k_i}h_m^2=-\frac{1}{L_i} \left(\int_{0}^{t_1}F(t) dt +\int_{t_2}^{t_f}F(t) dt \right),
\label{eq:E2}
\end{equation}
Defining $E_2=-\left(\int_{0}^{t_1}F(t) dt +\int_{t_2}^{t_f}F(t) dt \right)$ and eliminating $h_m$ to combine Eqs.~(\ref{eq:ice-melt}) and (\ref{eq:E2}), we have
\begin{equation}
    E_1 = E_2-\frac{B}{2k_iL_i}\cdot E_{1}^2.
\label{eq:limiting}
\end{equation}
where the left-hand side represents the energy used to melt the ice, while the right-hand side corresponds to the energy released due to ice growth. After assuming an eccentricity, we can use Kepler's equation to numerically solve Eq.~(\ref{eq:limiting}) for the mean stellar flux at which the Snowball-to-Waterbelt bifurcation occurs.

\begin{figure}
    \centering
    \includegraphics[width=0.5\linewidth]{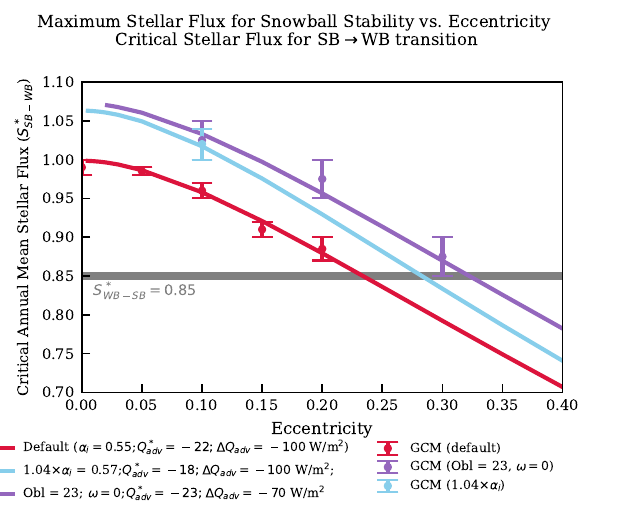}
    \caption{Maximum stellar flux allowing a Snowball state as a function of eccentricity. Red dots represent default GCM results, corresponding to the right boundary of the bistability region in Fig.~\ref{fig:bifur-diagram}. The error bars indicate the range in annual-mean stellar flux spanned by the two simulations closest to this boundary. We also plot GCM sensitivity tests with the ice albedo increased by a factor of 1.04 (blue dots) and the obliquity increased to 23 degrees (purple dots). The red line shows results from the ice-thermodynamic model with $\alpha_i$ and $Q_{adv}$ tuned to default GCM results at low eccentricity. The blue and purple lines show results of the ice-thermodynamic model with parameters adjusted for the GCM sensitivity tests. The horizontal gray line shows the approximate critical stellar flux for the transition from Waterbelt to Snowball in the GCM, corresponding to the left boundary of the bistability region.}
    \label{fig:threshold}
\end{figure}


Figure~\ref{fig:threshold} compares predictions for the critical annual mean stellar flux of the Snowball-to-Waterbelt bifurcation from the ice-thermodynamic model (Eq.~(\ref{eq:limiting})) with the GCM as a function of eccentricity. The ice-thermodynamic model with $Q_{adv}$ and $\alpha_i$ tuned closely reproduces the results of the GCM. In particular, it shows a similar pattern of the decrease in the annual mean stellar flux of the Snowball-to-Waterbelt bifurcation as eccentricity increases, with the slope becoming steeper as the eccentricity increases. This is important because we identified the decrease in $S^*_{SB\rightarrow WB}$ as eccentricity increases as the main reason the GCM loses Snowball climate bistability at increased eccentricity in section~\ref{sec:gcm-bistability}. The ice-thermodynamic model is also able to match GCM behavior, with slight retuning, when important variables are changed including the ice albedo ($\alpha_i$) and the obliquity. This is particularly impressive in the case of changing the obliquity because this changes both $E_1$ and $E_2$ in Equation~(\ref{eq:limiting}). As a result of the decrease in the annual mean stellar flux of the Snowball-to-Waterbelt bifurcation as the eccentricity increases, it eventually crosses the roughly constant annual mean stellar flux of the Waterbelt-to-Snowball bifurcation. The eccentricity at which this occurs ($e \approx 0.25$ in the default case) is the critical eccentricity at which Snowball climate bistability is lost. In both the GCM simulations and ice-thermodynamic model, we fix the orbital period at 360 days to isolate the effect of ice thermodynamics. Allowing the orbital period to vary according to Kepler's third law would lower the critical eccentricity slightly.

\subsection{Robustness of Results and Effect of Varying Parameters}\label{sec:robust}

\begin{figure}[hbt!]
    \centering
    \includegraphics[width=0.5\linewidth]{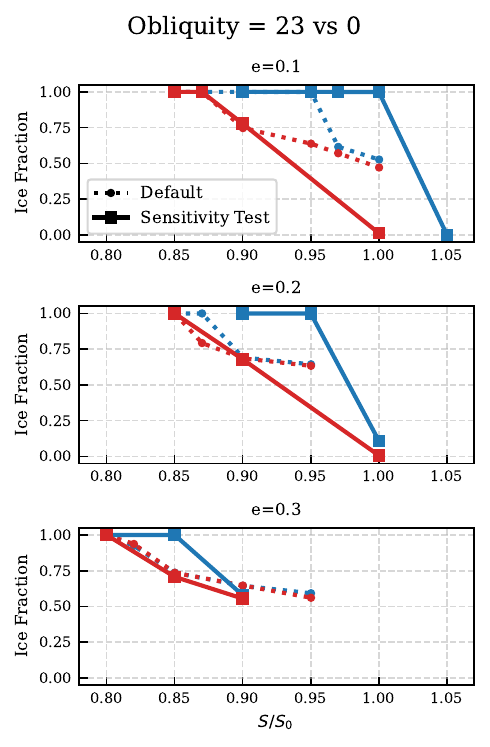}
    \caption{GCM bifurcation diagram for Earth-like obliquity cases. The format follows Fig.~\ref{fig:bifur-diagram}, where red markers represent Warm-Start simulations and blue markers represent Cold-Start simulations. Solid lines indicate simulations with Earth-like obliquity ($23^\circ$), while dotted lines show corresponding zero-obliquity cases for comparison. From top to bottom, eccentricity increases.}
    \label{fig:tests_obl}
\end{figure}

In our default simulations, we assume zero obliquity, resulting in a symmetric insolation pattern between the Northern and Southern Hemispheres throughout the year. Nonzero obliquity introduces a seasonal cycle in the latitudinal distribution of stellar flux, and the angle of periastron determines which hemisphere receives more annual-mean insolation. Changing obliquity alters the insolation pattern, temperature gradients, large-scale circulation, ice distribution, and other climate processes \citep[e.g.,][]{williams_habitable_1997, jenkins_global_2000, armstrong_effects_2014, spiegel_habitable_2009, ferreira_climate_2014, linsenmeier_climate_2015, rose_ice_2017,kilic_stable_2018,  kang_regime_2019, adams_aquaplanet_2019,  guendelman_atmospheric_2020, wilhelm_ice_2022, kodama_climate_2022, vervoort_system_2022, way_exploring_2023, hammond_coupled_2024}, which could potentially impact our results. To test this, we conducted simulations with an Earth-like obliquity of 23$^\circ$. We find that increasing the obliquity to 23$^\circ$ has little effect on the Waterbelt-to-Snowball transition and shifts the Snowball-to-Waterbelt transition to higher stellar flux (Fig.~\ref{fig:tests_obl}). As we increase the eccentricity, the bistability range decreases, and we expect it to disappear at an eccentricity slightly higher than 0.3  (Fig.~\ref{fig:tests_obl}). This behavior is fit well by the ice thermodynamic model with adjustments to the heat flux terms due to the change in the annual-mean stellar flux pattern (Fig.~\ref{fig:threshold}). Overall, we find very similar qualitative behavior for an obliquity of 23$^\circ$ as for an obliquity of 0$^\circ$.


\begin{figure}[hbt!]
   \centering
   \includegraphics[width=0.5\linewidth]{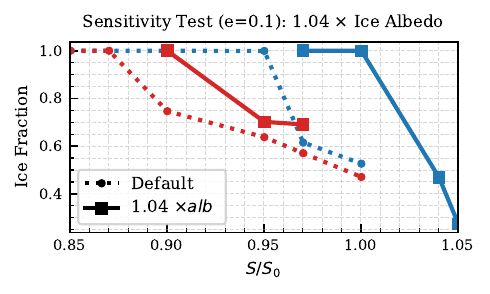}
   \caption{GCM bifurcation diagram with e=0.1, with the ice surface albedo increased by 4\%, from 0.67 to 0.7. Thick solid lines indicate simulations with the ice surface albedo adjusted, while dashed lines show the default case.}
   \label{fig:alb_tests}
\end{figure}

In Fig.~\ref{fig:alb_tests}, we show GCM simulations in which we increase the ice albedo by 4\%. Increasing the ice albedo increases both $S^*_{\text{SB} \rightarrow \text{WB}}$ and $S^*_{\text{WB} \rightarrow \text{SB}}$. However, the increase is more pronounced for $S^*_{\text{SB} \rightarrow \text{WB}}$, resulting in an expanded bistability region. This is reasonable, given that most of the tropics are ice-free in the Waterbelt. 


\begin{figure}
    \centering
    \includegraphics[width=0.4\linewidth]{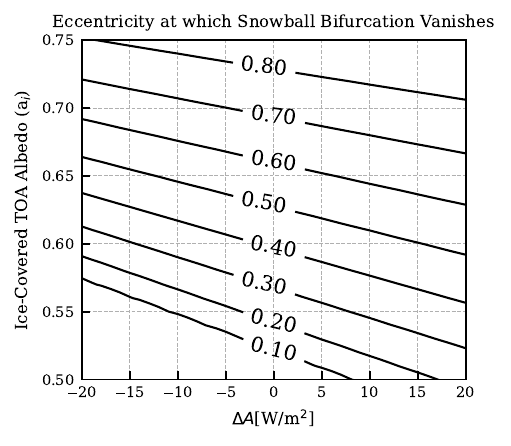}
    \caption{The maximum eccentricity at which a Snowball bifurcation can exist is shown as a function of the variation in infrared emission to space at the water freezing temperature, $\Delta A$, and top-of-atmosphere albedo over ice, $\alpha_i$, in the ice-thermodynamic model.}
    \label{fig:e_cri}
\end{figure}

Since we have established that the ice-thermodynamic model can quantitatively reproduce the GCM results, we can use it to investigate the effect of varying other parameters, which would be too expensive to do with GCM simulations. The most uncertain aspect of climate modeling is clouds, particularly in the Snowball climate \citep{abbot2012clouds,abbot2014resolved}. Clouds can both influence the planet's reflectivity ($\alpha_i$ in the ice-thermodynamic model) and infrared emission to space ($A$ in the ice-thermodynamic model). In Figure~\ref{fig:e_cri}, we vary these two parameters in the ice-thermodynamic model and study their effect on the critical eccentricity at which Snowball climate bistability is lost. Here we assume a fixed value of $S^*_{\text{WB} \rightarrow \text{SB}}=0.85\ S_0$ to focus on the uncertainty related to cloud behavior in the Snowball. If the ice-covered top-of-atmosphere albedo is higher due to more or brighter clouds, it becomes harder to get the Snowball temperature above freezing. As a result, the eccentricity must be increased more to cause the Snowball-to-Waterbelt transition, and the critical eccentricity increases (Fig. \ref{fig:e_cri}). Clouds are unlikely to increase the top-of-atmosphere albedo in the Snowball much \citep{abbot2012clouds,abbot2014resolved}, and other factors such as dust likely decrease the top-of-atmosphere albedo \citep{abbot2010mudball,abbot2010dust}, so that the large increases in $\alpha_i$ necessary to significantly increase the critical eccentricity are probably not realistic in most cases. The cloud effect on infrared emission is more uncertain \citep{abbot2012clouds,abbot2014resolved}, but it has a smaller effect on the critical eccentricity and is unlikely to change our qualitative story that Snowball climate bistability will be lost at a moderate eccentricity. Our sensitivity analysis with the ice-thermodynamic model therefore suggests that our main result is robust to plausible uncertainty in cloud modeling.


These results can be directly applied to other physical parameters. For example, the TOA albedo above ice could be up to 0.3 lower for planets orbiting M and K stars than G-stars due to differences in the stellar spectra \citep{shields_effect_2013, shields_spectrum-driven_2014}. This suggests that planets orbiting small stars will have an even smaller critical eccentricity at which Snowball climate bistability is lost, if it exists at all, since we do not expect Snowball climate bistability for tidally locked planets \citep{checlair_no_2017}. Also, including \ch{CO2} ice increases the TOA albedo, which could lead to a higher critical eccentricity \citep{Venkatesan_2025}. Additionally, advective heat flux, $Q_{adv}$, enters the model (Eq.~(\ref{eq:EBM})) the same way as $A$, but with the opposite sign. This means that a negative $\Delta A$ can be interpreted as equivalent to a positive $\Delta Q_{adv}$, which corresponds to less poleward heat transport given our sign convention ($Q_{\mathrm{adv}}<0$ for poleward flux). The advective heat flux should be inversely related to planetary rotation rate \citep{kaspi_atmospheric_2015, komacek_atmospheric_2019, williams_clouds_2024}, such that increasing the rotation rate would correspond to a positive $\Delta Q_{adv}$ or a negative $\Delta A$. Figure~\ref{fig:e_cri} therefore suggests that planets with a larger rotation rate than Earth ($\Delta A<0$) should tend to lose Snowball bistability at a smaller critical eccentricity, and those with a smaller rotation rate than Earth ($\Delta A>0$) should tend to lose Snowball bistability at a larger critical eccentricity. This makes sense because decreasing $A$ warms the planet and makes it easier to melt equatorial ice in a Snowball, all else being equal.

\begin{figure}
    \centering
    \includegraphics[width=0.5\linewidth]{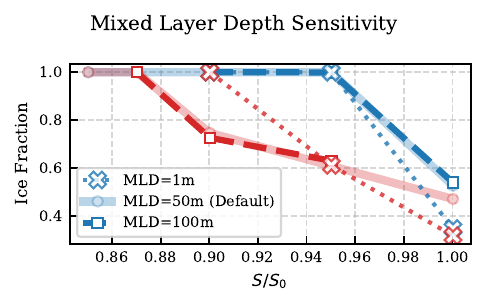}
    \caption{GCM bifurcation diagram with e=0.1, for mixed layer depth (MLD) values of to 1 m and 100 m in addition to the default value of 50 m.}
    \label{fig:MLD}
\end{figure}

We also tested the sensitivity of our results to the mixed-layer depth (MLD), which could vary with the planetary rotation rate.  The critical stellar flux for the Snowball-to-Waterbelt transition is not sensitive to varying the MLD from 1-m to 100-m (Fig. \ref{fig:MLD}). This outcome is consistent with our explanation of this transition, which focuses on ice thermodynamics such that ocean effects should be secondary. The Waterbelt-to-Snowball transition, however, does occur at lower eccentricity with a 1-m MLD due to enhanced seasonal variability, but does not change for 100-m MLD, as the seasonal variations are already effectively damped out for a MLD 50 m.

\begin{figure}
    \centering
    \includegraphics[width=0.4\linewidth]{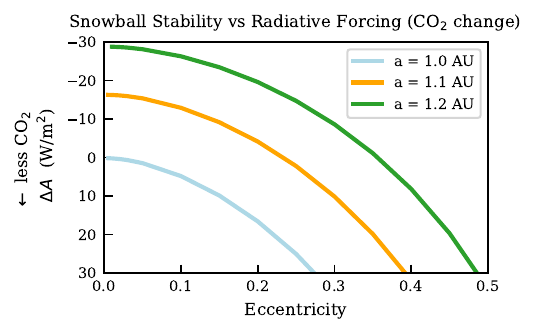}
    \caption{Prediction of maximum \ch{CO2} forcing allowing a Snowball state as a function of eccentricity using the ice-thermodynamic model. We explore the effect of changing \ch{CO2} concentrations by adjusting the parameter $A$ in the linearized OLR parameterization, where lower \ch{CO2} levels correspond to larger $A$ values due to weaker greenhouse effect.}
    \label{fig:CO2_forcing}
\end{figure}

In this study we have used the annual-mean stellar flux as the bifurcation parameter, but Snowball bifurcations can also be triggered by other forcings such as changes in atmospheric \ch{CO2} concentration. The key distinction between these two forcings is that changing the stellar flux modifies both the annual mean and the amplitude of seasonal variation in $F(t)$ (the purple curve in panel (a) of Fig. \ref{fig:EBM_GCM}), while varying \ch{CO2} only shifts the curve vertically. Here we fix the stellar flux and investigate the effect of \ch{CO_2} forcing. To simplify the analysis, we focus on changes in the longwave emission parameter $A$ to represent the greenhouse effect of \ch{CO2}, while keeping the radiative feedback parameter $B$ fixed. Consistent with when we vary the stellar flux, the critical \ch{CO2} concentration required to sustain a Snowball state decreases with increasing eccentricity (Fig.~\ref{fig:CO2_forcing}), such that Snowball bistability would disappear above a threshold eccentricity.

\section{Discussion}\label{sec:discuss}

Our proposed basic physical mechanism of ice self-insulation leading to ice thinning for increased seasonality is likely to be robust across different models. However, the relative importance of other factors may vary due to uncertain parameterizations of ice, snow, clouds, and other processes. For example, in our ExoCAM simulations, the equatorial albedo decreases to about 0.45 at perihelion and increases to about 0.65 at aphelion due to factors such as snow and ice melt, snow thickness, snow age, and cloud behavior. Despite this annual cycle in albedo, the annual-mean equatorial albedo is roughly independent of eccentricity in our simulations. As a result, it is possible for us to fit the ice-thermodynamic model to the GCM results using the annual-mean albedo. If the summer decrease in albedo were much lower in another GCM, this could lead to melting that might play a more prominent role in decreasing ice thickness with increased seasonality. Alternatively, a large winter increase in albedo could counteract the ice self-insulation effect, possibly even leading to an increase in ice thickness with increased seasonality in an extreme case, which would widen Snowball climate bistability at higher eccentricity. 

Both our GCM and ice-thermodynamic model neglect basal heat flux from below the ice. Without it, ice thickness could grow without bound at sufficiently low stellar flux. Even a small basal heat flux would cap the maximum ice thickness, but would have little influence on the thin-ice transition states considered here. Basal heat flux, if included, appears explicitly in the column energy budget (Eq. \ref{eq:EBM}), but only affects the surface temperature indirectly via $h$ in ice top energy balance of Eq. \ref{eq:Ts}, so its role differs from that of the heat-flux constant $A$. Including a 2 W/m$^{-2}$ basal heat flux changes the ice thickness by only $\sim$0.15 m in the example from Fig. \ref{fig:EBM_GCM}, leaving our conclusions unaffected. For exoplanets with moderate eccentricity, tidal deformation by the host star can convert orbital energy into heat. Using Eq. 2 of \citet{driscoll_tidal_2015}, we estimate an upper bound on tidal heating for the $S/S_0=1$, $e=0.2$ case, and find fluxes below 0.1 W m$^{-2}$ for stellar masses $>$0.27 $M_\odot$. Lower-mass stars are more likely to host tidally locked planets, which previous studies have shown lack Snowball bistability \citep{checlair_no_2017, checlair_no_2019}. Thus, tidal heating is unlikely to affect our conclusions.


Our work could be extended with a thorough investigation of the effects of land and ocean on our results. We assume an aquaplanet with no continents, but continents would decrease the planetary heat capacity and could increase climate sensitivity to eccentricity. We also do not consider the effect of a dynamical ocean and sea ice dynamics, which can strongly influence Snowball initiation \citep{voigt2012sea,rose2015stable}. Also, for a wide range of obliquities, planets may exhibit equatorial ice belts with open ocean at the poles \citep[e.g.,][]{jenkins_global_2000, rose_ice_2017, kilic_stable_2018, wilhelm_ice_2022}. A comprehensive investigation of how the combined effects of high obliquity and eccentricity influence stability of different ice configurations is beyond the scope of this study and warrants future investigation.

\section{Conclusion}\label{sec:conclusion}

This study explores Snowball climate bifurcations and bistability for terrestrial planets with varying orbital eccentricities using a Global Climate Model (GCM) and low-order ice thermodynamic model. Our main conclusions are as follows:

\begin{enumerate}
    \item Climate bistability between Snowball and Waterbelt states disappears at moderate eccentricity ($e > 0.25$ for zero-obliquity). This happens because the critical stellar flux for exiting the Snowball state ($S^*_{\text{SB} \rightarrow \text{WB}}$) decreases significantly with eccentricity, while the critical stellar flux for entering the Snowball state ($S^*_{\text{WB} \rightarrow \text{SB}}$) remains relatively constant. 

    \item The decrease in the critical stellar flux for exiting the Snowball state with eccentricity is driven by larger summer ice melting than winter ice freezing with increased seasonality. This occurs because ice freezing at the bottom (ocean-ice interface) is reduced by the insulation provided by the ice itself.

    \item The qualitative mechanism outlined here is not sensitive to reasonable variations in a number of variables, including planetary obliquity, ice albedo, and cloud radiative effect. 
\end{enumerate}

The finding that Snowball bistability vanishes at moderate eccentricity, together with previous results showing its absence for tidally locked planets, suggests that Earth-like Snowball episodes are unlikely in a number of exoplanet contexts, providing a potential datum in support of the Rare Earth Hypothesis \citep{ward_rare_2000}.

\begin{acknowledgments}
This work was supported by NASA award No.80NSSC21K1718, which is part of the Habitable Worlds program.  This work was completed with resources provided by the University of Chicago Research Computing Center. This research made use of the NASA Exoplanet Archive, which is operated by the California Institute of Technology, under contract with the National Aeronautics and Space Administration under the Exoplanet Exploration Program. We thank Katlin Hill, Mary Silber, Bowen Fan, Yaoxuan Zeng, Dan Fabrycky, Edwin Kite, Gongjie Li and Brandon Park Coy for comments. We thank Ian Eisenman and the anonymous reviewer for their detailed feedback and insightful suggestions, which significantly improved the clarity and quality of this manuscript. 
\end{acknowledgments}

\software{Numpy \citep{harris2020array}, Matplotlib \citep{Hunter:2007}}

\bibliographystyle{aasjournal}
\bibliography{snowball}

\end{CJK*}
\end{document}